# Experimental observation of energy-band Riemann surface

Dali Cheng,[1,2,*] Heming Wang,[1,2,*] Janet Zhong,[1,3]
Eran Lustig,[1] Charles Roques-Carmes,[1] and Shanhui Fan[1,2,3,#]

[1]*Ginzton Laboratory, Stanford University, Stanford, California 94305, USA*
[2]*Department of Electrical Engineering, Stanford University, Stanford, California 94305, USA*
[3]*Department of Applied Physics, Stanford University, Stanford, California 94305, USA*
[*]equal contributions, [#]shanhui@stanford.edu

**Abstract**

Non-Hermiticity naturally arises in many physical systems that exchange energy with their environment. The presence of non-Hermiticity leads to many novel topological physics phenomena and device applications. In the non-Hermitian energy band theory, the foundation of these physics and applications, both energies and wavevectors can take complex values. The energy bands thus become a Riemann surface, and such an energy-band Riemann surface underlies all the important signatures of non-Hermitian topological physics phenomena. Despite a long history and recent theoretical interests, the energy-band Riemann surface has not been experimentally studied. Here we provide a photonic observation of the energy-band Riemann surface of a non-Hermitian system. This is achieved by applying a tunable imaginary gauge transformation on the platform of the photonic synthetic frequency dimension. From the measured topology of the Riemann surface, we reveal the complex-energy winding, the open-boundary-condition spectrum, the generalized Brillouin zone, and the branch points. Our findings demonstrate a unified framework in the studies of diverse effects in non-Hermitian topological physics through an experimental observation of energy-band Riemann surfaces.

**Introduction**

A non-Hermitian system exchanges energy with its environment(*1–14*). In photonics, non-Hermiticity arises naturally because of gain and loss of the medium. Recent studies have found that the presence of non-Hermiticity leads to many topological physics phenomena without Hermitian counterparts, including the non-Hermitian skin effect(*15, 16*), exceptional points and exceptional rings(*7*), additional symmetry classes(*17*), with potential applications such as lasing(*18, 19*), sensing(*4, 20, 21*), and communications(*8, 22, 23*).

In the study of non-Hermitian energy band theory(*15, 17, 24–29*) that underlies these physics and applications, the system's energies and the wavevectors are generally complex. Thus, the band structure of a non-Hermitian system is a mapping between complex energies and complex wavevectors. Such a complex-valued mapping can be formulated as an energy-band Riemann surface(*30–32*). It has been theoretically shown that all important experimental signatures of the non-Hermitian energy band theory, including the spectrum for systems truncated with an open boundary condition (OBC)(*33–35*), the generalized Brillouin zone (GBZ)(*36, 37*), the winding(*38*) and braiding(*39*) of energy bands, and the branch points, are related to specific topological features on such energy-band Riemann surfaces(*30–32*). The concept of the energy-band Riemann surface in fact has a long history dating back to the pioneering work of Kohn in 1959(*40*). Despite this long history and recent theoretical interests, however, there has not been an experimental study of energy-band Riemann surfaces in any physical system.

In this work, we provide an experimental observation of energy-band Riemann surfaces in non-Hermitian photonic systems. We use a photonic ring resonator subject to periodic modulations as the experimental platform, and create non-Hermitian lattice models in the synthetic frequency dimension(*41–43*). We experimentally perform a tunable imaginary gauge transformation(*44*) for photons, which enables us to probe the system behavior at arbitrary complex wavevectors and to reconstruct the entire energy-band Riemann surface. From the experimentally measured energy-band Riemann surface, we determine important observables of the non-Hermitian lattice model, including the complex-energy windings, the OBC spectrum, the GBZ, and the branch points. Our results provide an experimental demonstration of the energy-band Riemann surface, a physical concept of long-standing theoretical interest, and a unified approach to the measurement of observables and signatures in non-Hermitian physics.

**Results**

**Measurement of energy-band Riemann surface**

We begin by considering a general one-dimensional tight-binding lattice model with translational invariance. The Hamiltonian is given by

$$H_0 = \sum_{x \in \mathbb{Z}} \sum_{n \in \mathbb{Z}, -r \leq n \leq r} g_n a_{x+n}^\dagger a_x$$

(1)

Here and throughout this paper, we take the lattice constant as 1, $x$ is the spatial coordinate of lattice sites, $\mathbb{Z}$ is the set of integers, $r$ is the maximum coupling range, $g_n$ is the coupling strength from site $x$ to site $x+n$ ($g_0$ is the on-site potential term), and $a_x$ and $a_x^\dagger$ are annihilation and creation operators on site $x$, respectively. Throughout the paper for simplicity we assume that both $g_r$ and $g_{-r}$ are non-zero. If the system is Hermitian ($H_0 = H_0^\dagger$), the coupling strengths satisfy $g_n^* = g_{-n}$; for general non-Hermitian models, $g_n$ can be arbitrary complex numbers.

The band structure connects the energy $E$ of a state with respect to its wavevector $k$. The band structure of $H_0$ is

$$E(k) = \sum_{-r \leq n \leq r} g_n e^{-ink} = g_0 + \sum_{1 \leq n \leq r} (g_n + g_{-n}) \cos nk + \sum_{1 \leq n \leq r} i(g_{-n} - g_n) \sin nk$$

(2)

Here $E$ is a general complex number, and $k$ is in the strip $0 \leq \mathrm{Re}(k) < 2\pi$, $-\infty < \mathrm{Im}(k) < +\infty$ on the complex-$k$ plane. The set of such $(E, k)$ pairs that satisfy the band structure form the energy-band Riemann surface.

To experimentally study the energy-band Riemann surface requires the probing of the system at arbitrary complex wavevectors. In many systems it is possible to probe the energy band structure at a real wavevector(*45–49*), which is the intersection of the plane $\mathrm{Im}(k) = 0$ and the energy-band Riemann surface and corresponds to the periodic-boundary-condition (PBC) spectrum. To probe the entire Riemann surface, we perform an imaginary gauge transformation on $H_0$, as represented by $k \to k + \mathrm{i}\sigma$, where $\sigma$ is real. The gauge-transformed band structure reads

$$E(k + \mathrm{i}\sigma) = \sum_{-r \leq n \leq r} g_n \mathrm{e}^{-\mathrm{i}n(k+\mathrm{i}\sigma)} = \sum_{-r \leq n \leq r} g_n \mathrm{e}^{n\sigma} \mathrm{e}^{-\mathrm{i}nk} \tag{3}$$

which corresponds to the lattice model

$$H(\sigma) = \sum_{x \in \mathbb{Z}} \sum_{n \in \mathbb{Z}, -r \leq n \leq r} (g_n \mathrm{e}^{n\sigma}) a^\dagger_{x+n} a_x \tag{4}$$

The PBC spectrum of $H(\sigma)$ gives the intersection of the plane $\mathrm{Im}(k) = \sigma$ and the energy-band Riemann surface of $H_0$. The entire energy-band Riemann surface of $H_0$ can therefore be reconstructed through a set of measurements at different values of $\sigma$.

We use the photonic synthetic frequency dimension(*41–43*) as the experimental platform because of its versatility and programmability. Fig. 1(a) is a sketch of our experimental setup; see Supplementary Materials for a detailed description. We consider a ring resonator that supports equally spaced frequency modes. The resonator is subject to periodic phase and amplitude modulations that couple neighboring frequency modes. If each frequency mode is viewed as a lattice site, the modulations will provide couplings between neighboring sites, and thus a tight-binding lattice model is formed. The resonator is excited by a continuous-wave laser of detuning $\delta\omega$ from one frequency mode, and the transmission intensity signal $\xi$ is measured as a function of $\delta\omega$ and modulation time $t$. Such transmission signal encodes information on the band structure of the tight-binding lattice model(*38, 39, 50*): in $\xi(\delta\omega, t)$, the time $t$ is interpreted as the wavevector $k$. At each wavevector $k$, the center location of the resonance corresponds to $\mathrm{Re}(E(k))$, and the linewidth corresponds to $\mathrm{Im}(E(k))$. The transmission measurement thus provides information of the band structure for real-valued wavevectors.

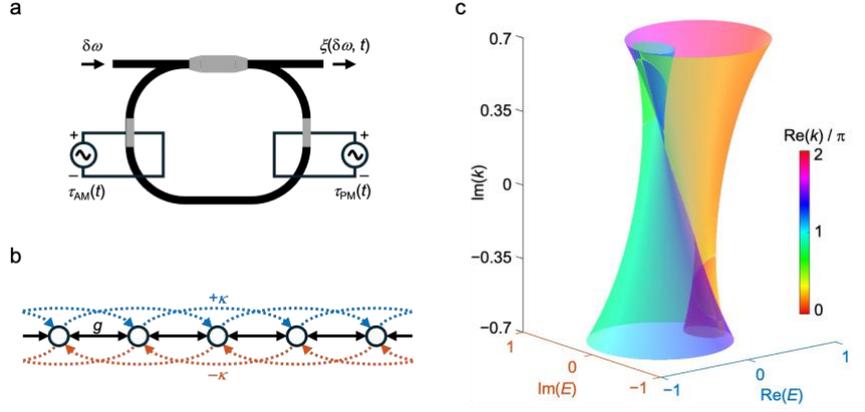

Fig. 1 Photonic measurement of the energy-band Riemann surface. (a) Sketch of the experimental setup, a ring resonator under phase modulation (PM) and amplitude modulation (AM). $\tau_{PM}(t)$ and $\tau_{AM}(t)$ represent the time-dependent transmittance of the modulators. The transmission light intensity $\xi$ is a function of the input frequency detuning $\delta\omega$ and the modulation time $t$. (b) Non-Hermitian tight-binding lattice model $H_1$. The nearest-neighbor coupling $g$ is symmetric, the next-nearest-neighbor coupling $\pm\kappa$ is anti-symmetric, and we take $g = 2\kappa = 0.2$ throughout the paper. (c) Energy-band Riemann surface of $H_1$. Re($k$) is encoded in the color of the sheets.

We combine the band structure measurement technique with the imaginary gauge transformation to measure the entire energy-band Riemann surface. As an example, we consider a lattice model with non-Hermitian next-nearest-neighbor couplings. Fig. 1(b) shows its lattice picture, and the Hamiltonian is

$$H_1 = \sum_x \left[ g(a^\dagger_{x+1} a_x + a^\dagger_{x-1} a_x) + \kappa(a^\dagger_{x+2} a_x - a^\dagger_{x-2} a_x) \right]$$

(5)

where $g = 2\kappa = 0.2$ are coupling strengths. In the form of Eq. (1), we take $g_1 = g_{-1} = g$, $g_2 = \kappa$, $g_{-2} = -\kappa$, and $g_n = 0$ for any other $n$. The energy band of $H_1$ is $E(k) = 2g\cos k - 2i\kappa\sin 2k$, and Fig. 1(c) plots the energy-band Riemann surface in the (Re($E$), Im($E$), Im($k$)) space, with colors of the sheets representing Re($k$) information. The energy-band Riemann surface of the Hermitian model with $\kappa = 0$ is shown in Fig. S4.

Under the weak-modulation approximation(51), $H_1$ can be implemented in the synthetic frequency dimension by modulation signals $\tau_{PM}(t) = e^{-2ig T_R \cos\Omega_R t}$ and $\tau_{AM}(t) = e^{-2\kappa T_R \sin 2\Omega_R t}$, where $\tau(t)$ is the time-dependent transmittance of the modulator, PM refers to phase modulation, and AM to amplitude modulation(38). $\Omega_R$ is the free spectral range of the resonator, and $T_R = 2\pi/\Omega_R$ is the round-trip time. The imaginary gauge transformation can be achieved by the substitution in the modulation signals: $\Omega_R t \to \Omega_R t + i\sigma$ (see Supplementary Materials). With this substitution, the modulation signals become

$$\tau_{\text{PM}}(t) = e^{-i(2gT_R \cosh\sigma \cos\Omega_R t + 2\kappa T_R \sinh 2\sigma \cos 2\Omega_R t)}$$

$$\tau_{\text{AM}}(t) = e^{-(2gT_R \sinh\sigma \sin\Omega_R t + 2\kappa T_R \cosh 2\sigma \sin 2\Omega_R t)}$$

(6)

Under modulation signals Eq. (6), the measured complex band energies correspond to the intersection of the energy-band Riemann surface and the plane $\text{Im}(k) = \sigma$. Note that the energy band of $H_1$ has symmetry

$$E(\pi - k) = -E(k)$$

(7)

and therefore it is sufficient in our experiments to focus on $\text{Im}(k) \geq 0$.

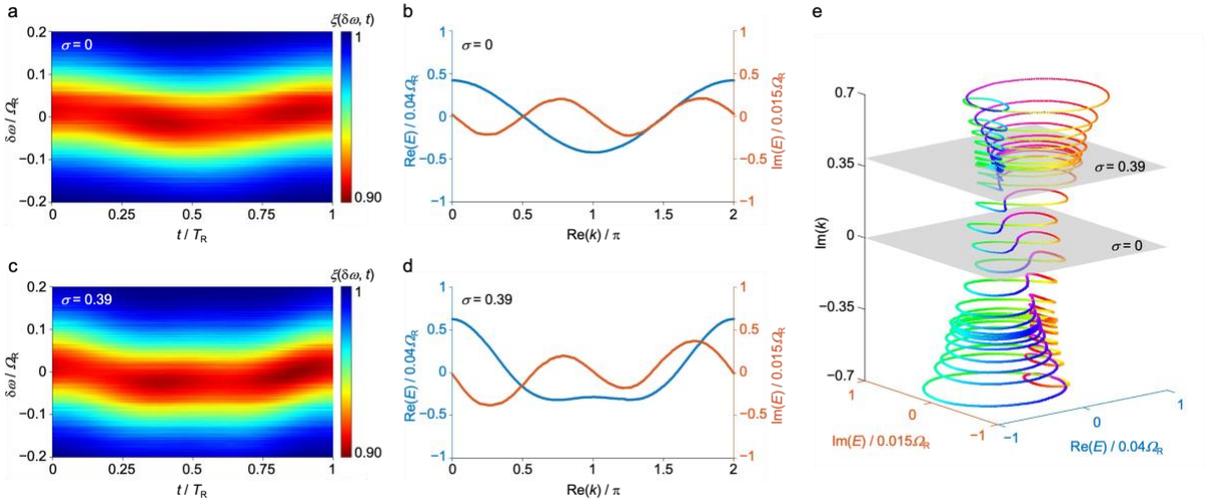

Fig. 2 Energy-band Riemann surface measurement of $H_1$ using imaginary gauge transformations. (a, c) Normalized transmission signal $\xi(\delta\omega, t)$ of the ring resonator. (b, d) Real and imaginary parts of the band energies as a function of $\text{Re}(k)$, extracted from the locations and linewidths of the resonance dips in (a, c), respectively. In (a, b), $\text{Im}(k) = \sigma = 0$; in (c, d), $\text{Im}(k) = \sigma = 0.39$. (e) Experimental measurement of the Riemann surface, by stacking the complex energies on different iso-$\text{Im}(k)$ planes. The grey planes highlight the locations of the data in (a, b) and (c, d).

Fig. 2 exhibits the experimental results of the energy-band Riemann surface measurement of $H_1$. In Figs. 2(a, b), we take $\sigma = 0$ in the modulation signal, without the imaginary gauge transformation. Fig. 2(a) shows the transmission signal $\xi(\delta\omega, t)$. A dip in the signal is clearly observed and corresponds to the resonance of the energy band. Fig. 2(b) shows the real and imaginary parts of the band energies as a function of $\text{Re}(k)$, by fitting the data in Fig. 2(a) with a Lorentzian lineshape (see Supplementary Materials for methods). These complex band energies, when $\text{Re}(k)$ is viewed as a parameter, compose the intersection of the Riemann surface and the plane $\text{Im}(k) = 0$ in the ($\text{Re}(E)$, $\text{Im}(E)$, $\text{Im}(k)$) space. In Figs. 3(c, d), we take $\sigma = 0.39$, with the imaginary gauge transformation. The complex band energies in Fig. 3(d) compose the intersection of the Riemann surface and the plane $\text{Im}(k) = 0.39$. In Fig. 3(e), we plot the entire energy-band Riemann surface, by stacking the complex energies obtained from different imaginary gauge transformations on their corresponding iso-Im(k) planes. Measurements of $\sigma = 0$ and $\sigma = 0.39$ are highlighted by grey planes. The data for $\sigma < 0$ are obtained from their $\sigma > 0$ counterparts by applying the symmetry relation of Eq. (7). The experimental results are in good agreement with the theoretical ones in Fig. 1(c).

The topology of a Riemann surface is characterized by its branch points and branch cuts. The band structure of the Hamiltonian in Eq. (1) has the form $E = \sum_{-r \leq n \leq r} g_n z^{-n}$, where $z = e^{ik}$. Given each $E$, there are $2r$ solutions of $z$ and hence $2r$ wavevector $k$'s within the strip $0 \leq \text{Re}(k) < 2\pi$. These wavevectors are labelled as $k_j$ with $j = 1, \ldots, 2r$, sorted by their imaginary parts in the non-decreasing order. The Riemann surface thus consists of $2r$ energy Riemann sheets. At certain energies, some wavevectors (typically two of them) can become degenerate (i.e., $k_j = k_{j+1}$ for a certain $j$). Such energies are the branch points of the Riemann surface and the corresponding degenerate wavevectors are the ramification points(52). These branch points are connected by branch cuts, defined as arcs on the energy Riemann sheets where $\text{Im}(k_j) = \text{Im}(k_{j+1})$. Different Riemann sheets are connected together through the branch cuts. As the energy goes around a branch point on a Riemann sheet, it passes through a branch cut and emerges on a different Riemann sheet. Below, we experimentally demonstrate that both the branch cuts and the branch points on the energy-band Riemann surface are associated with important physical properties of non-Hermitian systems.

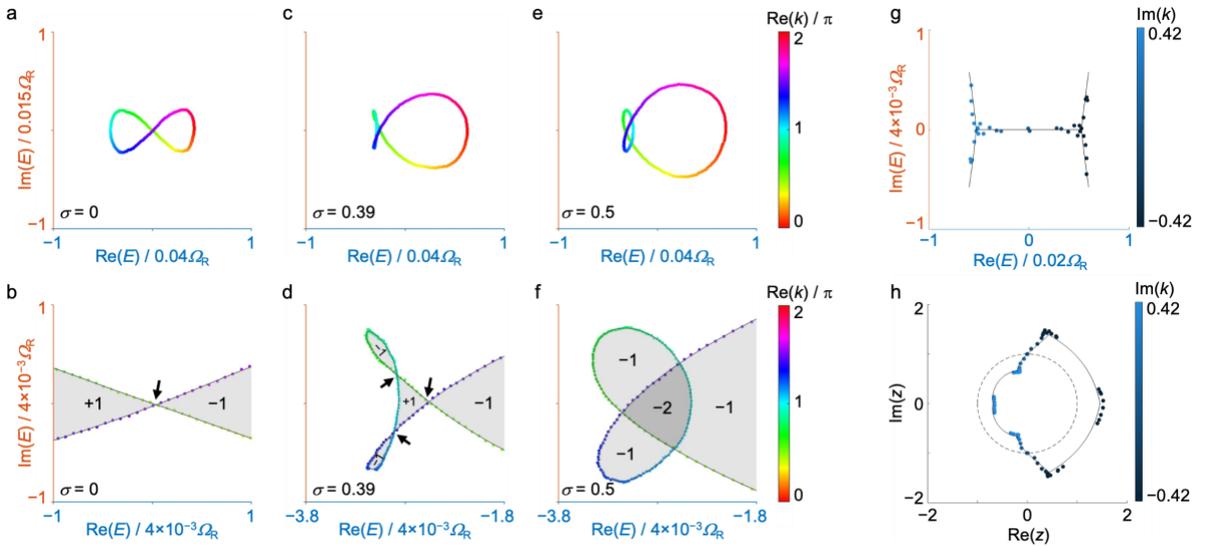

Fig. 3 Complex-energy windings, the OBC spectrum, and the GBZ of $H_1$ from its energy-band Riemann surface. (a–f) Complex-energy windings at (a, b) $\sigma = 0$, (c, d) $\sigma = 0.39$, (e, f) $\sigma = 0.5$. Dots represent experimental data of band energies whose colors encode Re(k). (b, d, f) are zoom-in plots of the complex-energy windings, with a focus on self-intersecting features and regions of different winding numbers. The black solid line is the periodic smoothing cubic splines fitting of the experimental data(53) (see Supplementary Materials). Regions of non-zero winding numbers are shaded in grey. The self-

intersecting points in the windings in (b, d) are associated with the OBC spectrum and the GBZ (see Supplementary Materials) and highlighted by black arrows, while those in (f) are not. (g) The OBC spectrum of $H_1$. (h) The GBZ of $H_1$ on the complex-$z$ plane, $z = e^{ik}$. In (g) and (h), dots represent experimental data, and the black solid line is the theoretical result. In (g), theoretical results of Re($E$) and Im($E$) are re-scaled by the ratios between the axes of Fig. 2(e) and Fig. 1(c). In (h) the dashed line represents the unit circle, i.e., the conventional Brillouin zone.

**OBC spectrum and GBZ as branch cuts on the Riemann surface**

For a non-Hermitian system, its OBC spectrum, i.e., the energy spectrum when the system is truncated with the open boundary condition, is drastically different from the PBC spectrum. Significant theoretical efforts have been carried out for the determination and understanding of the OBC spectrum(15, 17, 24, 26–29, 44, 54–58). The OBC spectrum is closely related to the concept of the GBZ. As mentioned above, for the Hamiltonian of Eq. (1), at each complex energy $E$, there are $2r$ possible wavevectors, $k_1, k_2, …, k_{2r}$, sorted by their imaginary parts in the non-decreasing order. The complex energies $E$ that satisfy Im($k_r$) = Im($k_{r+1}$) form the OBC spectrum, and the corresponding $k_r$'s and $k_{r+1}$'s form the GBZ(28). Thus, the OBC spectrum and the GBZ manifest as the branch cut between the $r$-th and the ($r+1$)-th Riemann sheets when sorted by Im($k$)(31, 32).

Here we demonstrate that our experimentally measured energy-band Riemann surface, constructed from the measurement of a set of PBC spectra, can be used to determine both the OBC spectrum and the GBZ. The intersection of a given slice of Im($k$) = $\sigma$ with the energy-band Riemann surface can be represented by the complex-energy winding, i.e., the winding of the trajectory of the complex energies as the Re($k$) varies from 0 to $2\pi$. The complex-energy winding at Im($k$) = 0 is the PBC spectrum and has been measured(38). In Figs. 3(a–f), we show the experimentally measured complex-energy windings at various values of $\sigma$. As $\sigma$ varies, we observe the variations of the winding patterns and the point-gap topologies, which are characterized by the winding numbers of the complex-energy trajectories with respect to different reference points on the complex-energy plane.

At a self-intersecting point in the complex-energy winding for a given slice of Im($k$) = $\sigma$, there are two different wavevectors $k$ and $k$' with the same complex energy and with Im($k$) = Im($k$') = $\sigma$. Such a self-intersecting point thus lies on a branch cut. To select the OBC spectrum among all branch cuts, instead of identifying the individual sheets of the Riemann surface, we can consider the winding numbers with respect to reference points in the regions surrounding the self-intersection points. A self-intersecting point is on the OBC spectrum if and only if it is surrounded by four regions of winding numbers 0, +1, 0, −1(32, 59) (see Supplementary Materials). According to this criteria, in Figs. 3(b, d), the self-intersecting points highlighted by the black arrows are on the OBC spectrum. In contrast, all the three self-intersecting points in Fig. 3(f), with $\sigma$ = 0.5, have a neighbouring region of winding number −2, and therefore none of them are on the OBC spectrum. Fig. 3(g) and Fig. 3(h) show the OBC spectrum and the GBZ of $H_1$, as determined by analyzing the experimetally obtained energy-band Riemann surface. The experimental data of the OBC spectrum with $Re(E) > 0$, and those of the GBZ outside the unit circle, are associated with $\sigma < 0$, and therefore they are obtained from the $\sigma > 0$ counterparts by applying the symmetry Eq. (7). Agreement is found between experimental and theoretical results.

Previously, both the OBC spectrum and the GBZ have been measured directly by constructing an open chain of electric(33, 36, 60), mechanic(34, 37), or photonic(35) resonances under the open boundary condition. In contrast, in our scheme, the OBC spectrum and the GBZ are measured in the bulk. Our work provides a direct demonstration of the connection between the OBC spectrum, the GBZ, and the topology of the energy-band Riemann surface, and highlights the topological nature of the OBC modes.

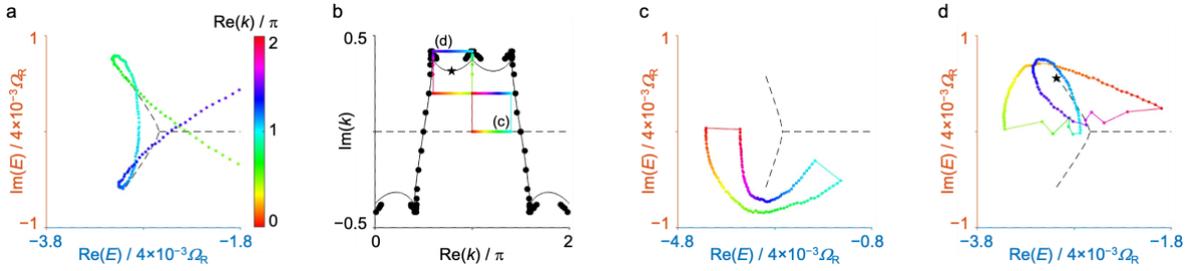

Fig. 4 Branch points of $H_1$ from its energy-band Riemann surface. (a) Complex-energy winding of $H_1$ at $\sigma = 0.38$. Colored dots represent experimental data. (b) GBZ of $H_1$ on the complex-$k$ plane. Black dots are experimental data, the black solid line represents theoretical results, and the black dashed line represents the conventional Brillouin zone $Im(k) = 0$. The colored dots represent two simple-loop trajectories on the complex-$k$ plane. One trajectory, labeled by (d), encircles a ramification point (black star), and the other, labeled by (c), does not. (c, d) Experimentally measured energies on the complex-$k$ trajectories in (b). The complex energies form a simple-loop trajectory in (c), and encircle the branch point twice in (d). In (a, c, d), black dashed lines represent the theoretical OBC spectra, re-scaled by the ratios between the axes of Fig. 2(e) and Fig. 1(c). In (b, d), the black stars mark the branch point of interest.

**Branch points on the Riemann surface**

Branch points are another topological feature of the energy-band Riemann surface with many physical implications. As mentioned above, a branch point corresponds to a specific complex energy $E$ where the wavevectors become degenerate. The energies of the branch points are given by the criterion $\partial E/\partial k = 0$. The termination points of the OBC spectrum are all branch points(32). The complex-energy winding features cusps at the energies of the branch points(61). The branch points are therefore a form of Van Hove singularities(62), where the density of states diverges(63). The branch points are also associated with saddle points(64), which directly influence the transient dynamics of waves propagating on the chain(*61*, *65*, *66*). Despite their important physical implications, direct measurements of branch points and their associated topological features have not been previously reported in the context of non-Hermitian band structures.

The model $H_1$ features four branch points located at $E = \pm 0.298 \pm 0.155i$, associated with the four termination points of the OBC spectrum in Fig. 3(g). Fig. 4(a) shows the complex-energy winding at $\text{Im}(k) = \sigma = 0.38$. For this choice of the imaginary gauge parameter $\sigma$, the branch point at $E = -0.298 + 0.155i$ is located near the complex-energy winding. In the vicinity of the branch point, the complex-energy winding encloses a small region with winding number $-1$ that resembles a cusp structure, which is an experimental signature of the branch point. As $\text{Im}(k)$ decreases, this region shrinks and collapses to the branch point; see Supplementary Materials for more discussions.

A branch point on the complex-$E$ plane corresponds to a ramification point on the complex-$k$ plane. Fig. 4(b) shows the GBZ in the complex-$k$ plane with the corresponding ramification point marked. In general, the ramification point corresponds to a local extremum of $\text{Im}(k)$ in such a plot of GBZ in the complex-$k$ plane; see Supplementary Materials for more discussions.

The topological features of branch points can be demonstrated by continuously moving the state on the energy-band Riemann surface. For a simple-loop trajectory on the complex-$k$ plane that does not enclose a ramification point, an example of which is illustrated in Fig. 4(b), the energy of the state also forms a simple loop, as shown in Fig. 4(c). In this case, there is a one-to-one mapping between the complex wavevector and the complex energy on their respective trajectories. On the other hand, for a simple-loop trajectory that encircles a simple ramification point on the complex-$k$ plane, an example of which is also illustrated in Fig. 4(b) for the ramification point mentioned above, the corresponding complex energy winds twice around the corresponding branch point as shown in Fig. 4(d). Our results here demonstrate the non-trivial topology of the energy-band Riemann surface near the branch point, and highlight the connection beween the branch point and other important aspects of non-Hermitian energy band theory including the OBC spectrum and the GBZ.

**Conclusions**

In this paper, we provide an experimental observation of the energy-band Riemann surface of a non-Hermitian photonic system. Using a periodically modulated ring resonator, the energy-band Riemann surface is measured by implementing a tunable imaginary gauge transformation for photons in the frequency domain. Our results demonstrate that the topology of the energy-band Riemann surface, controlled by its branch cuts and branch points, underlies important physical properties of non-Hermitian systems, including the complex-energy windings, the OBC spectrum, and the GBZ.

The scheme presented in this paper can also be generalized to other non-Hermitian photonic systems. Within tight-binding lattice models, it is possible to incorporate additional modulation frequencies in the resonator(*67–70*), and other degrees of freedom of photons including polarization(*71*, *72*), orbital angular momentum(*73*), and arrival time(*16*), to study energy-band Riemann surfaces in multi-dimensional, multi-band lattice models(*74, 75*). Beyond tight-binding lattice models, the energy-band Riemann surfaces of continuous photonic systems such as photonic crystals and metasurfaces can be studied through a transient (complex-energy)(*76*), evanescent (complex-wavevector)(*77*) excitation. Such continuous photonic systems can exhibit an additional richness of non-Hermitian topological phenomena(*78*).

**Acknowledgements**

The research is supported by a MURI project from the U.S. Air Force Office of Scientific Research (Grant No. FA9550-22-1-0339). We thank Prof. David A. B. Miller for providing laboratory space and equipment. C. R.-C. is supported by a Stanford Science Fellowship.

**Author contributions**

D. C. conceived the original idea. D. C. and H. W. did the experiment with contributions from E. L. and C. R.-C.. D. C., H. W., and J. Z. performed theoretical analysis. S. F. supervised the research. D. C., H. W., and S. F. wrote the manuscript with input from all authors.